\documentclass[reprint,amsmath,amssymb,braket,aps,prx,longbibliography]{revtex4-1} 
\usepackage{amsmath,amssymb}
\usepackage{graphicx,color,braket,amsmath,dcolumn,bm}
\usepackage[colorlinks=true,linkcolor=blue]{hyperref}

\begin{document}
\title{One-dimensional electron localization in semiconductors coupled to electromagnetic cavities}

\author{Dmitry Svintsov}
\email{svintcov.da@mipt.ru}
\affiliation{Moscow Institute of Physics and Technology, Dolgoprudny 141700, Russia}
\author{Georgy Alymov}
\affiliation{Moscow Institute of Physics and Technology, Dolgoprudny 141700, Russia}
\author{Zhanna Devizorova}
\affiliation{Moscow Institute of Physics and Technology, Dolgoprudny 141700, Russia}
\author{Luis Martin-Moreno}
\affiliation{Instituto de Nanociencia y Materiales de Aragon (INMA), CSIC-Universidad de Zaragoza, Zaragoza 50009, Spain}
\affiliation{Departamento de Fisica de la Materia Condensada, Universidad de Zaragoza, Zaragoza 50009, Spain}

\begin{abstract}
Electrical conductivity of one-dimensional disordered solids decays exponentially with their length, which is a celebrated manifeststation of localization phenomenon. Here, we study the modifications of localized conductivity induced by placement of 1d semiconductors inside of single-mode electromagnetic cavities, focusing on the regime of non-degenerate doping. We use the Green's function technique modified for non-perturbative account of cavity excited states, thus accounting for both coherent electron-cavity effects (i.e. electron motion in the zero-point fluctuating field) and inchoherent processes of photon emission upon tunneling. The energy spectrum of electron transmission in the cavity acquires Fano-type resonances associated with virtual photon emission, passage along the resonant level, and re-absorption. The quality factor of the Fano resonance depends on whether the intermediate state is coupled to the leads, and reaches its maximum when the intermediate state is localized deep in the disorder potential. Coupling to cavities also elevates the energies of shallow bound states, bringing them to the conduction band bottom. Such an effect leads to enhancement of low-temperature conductance.

\end{abstract}
\maketitle

\section{Introduction.}
\label{sec-Intro}
Manifestations of non-empty electromagnetic vacuum in real-life observables have been in the focus of research since the dawn of quantum field theory~\cite{Schwinger,Lamb,Casimir}. Most of manifestations, like anomalous magnetic moment of electron and Lamb shift, have been limited to atomic physics. Recent advances in fabrication of high-quality electromagnetic cavities~\cite{Strong_coupling_regimes} along with discoveries of ultra-confined electromagnetic waves~\cite{Koppens_confinement_limits,Nanotube_plasmons} have led to the idea of changing the macroscopic properties of matter with zero-point oscillations~\cite{FE_cavities_theory_1,SC_cavities_theory1,SC_cavities_theory_2,Hubbard_cavities_theory,Kibis_gap_by_vacuum,Enhanced_magnetism_in_QED,Polini_SC_cavities}. Remarkable experimental examples are manipulating the superconducting critical temperatures with cavities~\cite{SC_strong_coupling-experiment} and altering the rates of chemical reactions~\cite{Chem_strong_coupling}. At the same time, changing the solids' conductivity, the property of main interest in electronics, was long considered impossible even under resonant coupling~\cite{Limbacher2020}. Indeed, the momentum acquired by an electron in zero-point vacuum field is much less than thermal or Fermi momentum at most available temperatures and carrier densities.

Recent experiments have doubted this viewpoint. Ref.~[\onlinecite{Orgiu2015}] has shown the enhancement of conductivity in a disordered organic semiconductor proximate to a photonic crystal. Ref.~[\onlinecite{Faist_magneto_transport}] has demonstrated the effect of microwave cavities on the magnetoresistance of quantum wells. Further studies of similar systems demonstrated a finite dissipative conductivity on the fractional quantum Hall plateaus~\cite{Faist_QH_breakdown}. Ref.~\cite{Pisani2023} has shown the existence of strong photocurrent in the level anticrossing gap of the optical cavity. The complexity of these experiments lies the in comparison of the samples' properties with an without a cavity. This can be hindered by non-equivalent disorder, built-in fields, and geometric variations. Despite this complexity, it is intriguing to reveal theoretically the conditions under which the conductivity of solids may vary due to electromagnetic cavities. Once these conditions are met, introduction of cavities may become a new 'turning knob' for the manipulation of electrical properties, complementary to the doping and field effects.

Here, we show that an appropriate playground for pronounced manifestations of zero-point electromagnetic oscillations even at weak coupling is represented by one-dimensional (1d) disordered conductors in the localized regime~\cite{Localisation_exponent,Dorokhov_localization}. 
Related previous works considered mainly the transport in cavity-coupled ordered systems~\cite{Ciuti_vaccum_dressed_magnetotransport,Ground_state_electroluminescence,Hagenmuller_PRB,Hagenmuller_PRL,Eckhardt_exctly_solvable_TB_chain,Feiginov_RTDs_cavities}, or cavity modifications of individual scattering events~\cite{Iorsh_Scattering_in_cavity,Flambaum_tunneling}. In electrically uncoupled 1d systems, the action of cavity field on localization length was shown to be oscillatory at strong coupling~\cite{ge2022cavity}. Recently, refs.~\cite{Ciuti_2022,Ciuti_quantum_Hall} presented the studies of conductance and quantum Hall effect in cavity-coupled disordered 1d and 2d systems for in the case of degenerate doping.  They argued on emegence of cavity-induced long-range electron hopping. The complexity of this regime stands from the Pauli blocking of electronic states after emission of cavity photon. Such effects can be accurately accounted for only within the Keldysh Green's function technique. The latter has indeed been applied to related systems, the molecular transistors strongly coupled to phononic vibrations`\cite{Chen_STM,Arseev_2010}. Unfortunately, extension of this technique toward ultrastrong coupling is challenging due to complexity of diagrammatic expressions for electron-photon self-energies.
\begin{figure*}[ht!] 
    \includegraphics[width=1.0\linewidth]{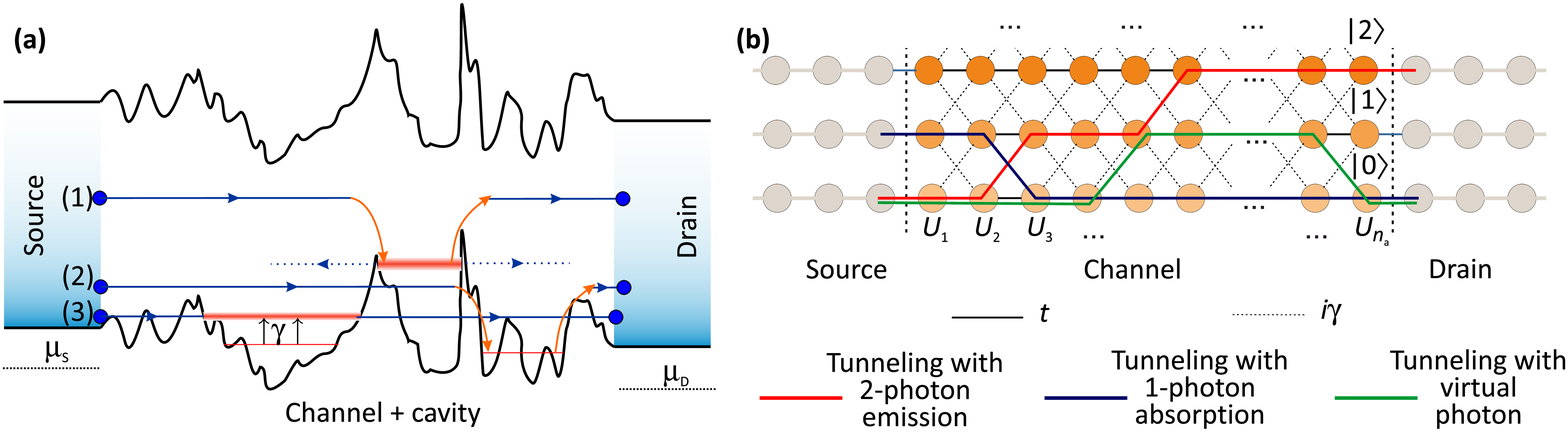}
    \caption{{\bf One-dimensional disordered semiconductor in a cavity.} (a) Mechanisms of conduction modifications via electromagnetic cavity fluctuations (1) An electron can emit a virtual photon, travel along an electronic quasi-bound resonant level, and absorb the photon back. The quasi-bound level is prone to decay into source and drain, which leads to the electroluminescence (2) Same as (1), but an electron is truly bound in the disorder potential, and no luminescence occurs (3) Coupling of conductor to a cavity with strength $\gamma$ leads to the 'expulsion' of the shallow bound states from the disorder potential. As a result, new resonances can appear in the transmittance (b) Extended states' space for a tight-binding chain strongly coupled to a cavity mode. Each atomic site with potential $U_i$ is now characterized by a number of cavity photons $\ket{N}$, and so do the states in the leads. Paths of different color show several possibilities of source-drain tunneling. Green -- elastic tunneling perturbed by emission/absorption of virtual photons, blue -- tunneling with single photon absorption, red -- tunneling with two-photon emission.
    }
    \label{fig1}
\end{figure*}

In this work, we focus on the conductivity of {\it non-degenerate} disordered 1d semiconductors coupled to single-mode cavities. The low fermionic occupation numbers $n_F \ll 1$ enable us to ignore the Pauli blocking effects and account for electron-cavity couplings numerically and non-perturbatively. Absence of Pauli blocking fully uncovers the strong coupling effects, primarily -- but not exclusively -- being the emissions of virtual photons upon transport. From experimental viewpoint, the non-degenerate regime is realized in intrinsic and moderately doped semiconductors, and is always achievable by gating. 

We find that coupling of a one-dimensional semiconductor to a cavity leads to the emergence of resonances in electron transmission $\mathcal{T}(E)$ at energies corresponding to the multi-photon replicas of quasi-bound states, $E=E_i + N \hbar \omega$ [process \# 1 in Fig.~\ref{fig1} (a)]. The replica resonances in transmission have a Fano-type structure. The probabilities of photon-assisted tunneling have resonances at the same replica energies, albeit having normal Lorentzian shape. At large photon energies $\hbar\omega \gg kT$, the emergence of such resonances does not lead to any modifications of one-dimensional conductivity. The reason is that such resonances appear at the high energy 'tail' of electron Boltzmann distribution. The situation changes at lower cavity photon energies $\hbar\omega \lesssim kT$ or at relatively large disorder amplitudes $\sigma \gtrsim \hbar \omega$. In such a case, replica resonances may appear in the vicinity of the nominal conduction band edge $E_C$ and greatly contribute to the enhanced conduction. The mechanism of conductivity in this regime represents electron hopping to the true bound state in the disorder potential with virtual photon emission, passage along this state, and subsequent photon absorption [process \# 2 in Fig.~\ref{fig1} (a)]. Finally, we find that coupling of disordered 1d wires to cavities can result in elevation of the bound states from disordered tail to the band, accessible from the leads [process \# 3 in Fig.~\ref{fig1} (a)]. At a coupling strength corresponding to the level unbinding, the thermally-averaged transmittance acquires a resonance. 

Our study is based on a modified Landauer-Buttiker formalism in the tight-binding approximation. The non-trivial modification involves the inclusion of cavity degrees of freedom populated according to the thermal Gibbs distribution. The developed method allows the evaluation of both direct tunneling modified by the zero-point oscillations of the cavity field and the tunneling aided by absorption/emission of real photons. Some examples of the relevant processes are shown in Fig.~\ref{fig1} (b). In Section \ref{sec-Landauer} we will describe the developed theoretical method. Section \ref{sec-Results} will discuss the results of this method applied to the conductance of disordered 1d chains coupled to the cavities. The final Section \ref{sec-discussion} discusses the possible experimental setups where the cavity modifications of conductance can be observed.

\section{Model: Landauer-Buttiker approach for a tight-binding chain in a cavity}
\label{sec-Landauer}
\subsection{Hamiltonian of the interacting chain}

The model system under consideration is the one-dimensional atomic chain with $n_{\rm at}$ sites. The electron transport in the chain is considered within the tight-binding approach, with constant nearest-neighbor hopping integral $t$ and random on-site energies $U_{i}$ (Fig.~\ref{fig1}). The amplitudes $U_i$ belong to the uncorrelated Gaussian distribution with variance $\delta U$.

The electromagnetic field of single-mode cavity is treated with Pierles modification of the hopping integrals~\cite{Eckhardt_exctly_solvable_TB_chain}, 
\begin{equation}
    t \rightarrow t \exp\{ \pm  i e A_x a / \hbar c\},
\end{equation}
where $A_x$ is the vector-potential along the chain and $a$ is the lattice constant. We restrict ourselves to the chains with length well below the radiation wavelength. In this limit, the quantization rule for the cavity field is simply $ A_x \rightarrow  A_{\rm vac} (\hat{b} + \hat{b}^+)$, where $\hat{b}^+$ and $\hat{b}$  are raising and lowering operators for the field, and
\begin{equation}
     A_{\rm vac} = \left(\frac{2\pi \hbar c^2}{\omega V}\right)^{1/2}
\end{equation}
is the amplitude of zero-point oscillations in the cavity of volume $V$. These prerequisites lead us to the following Hamiltonian $ {\hat{\mathcal H}}$, comprised of a free-field part $ {\hat{\mathcal H}}_{\rm f}$ and a part for the interacting chain $ {\hat{\mathcal H}}_{\rm ch}(\gamma)$:
\begin{gather}
\label{eq-Hamiltonian}
  {\hat{\mathcal H}} = \hat{\mathcal H}_{\rm f} + \hat{\mathcal H}_{\rm ch}(\gamma),\\
  \hat{\mathcal H}_{\rm f} = \hbar \omega {{\hat{b}}^{+}}\hat{b},\\
  \hat{\mathcal H}_{\rm ch}(\gamma) = \sum\limits_{i=1,\pm}^{n_A}{ U_i \hat{a}_{i}^{+}{{{\hat{a}}}_i}} +
  t e^{ \pm \frac{ i \gamma }{t}( \hat{b}+{{{\hat{b}}}^{+}} ) }
  \hat{a}_{i\pm 1}^{+}{\hat{a}_i}.
\end{gather}

Above, $\hat{a}_i$, $\hat{a}^+_i$ are fermionic annihilation and creation operators at site $i$, $\gamma = e a t A_{\rm vac} / \hbar c$ is the hopping amplitude associated with photon emission/absorption, and $n_A$ is the number of atoms in a chain. 

\subsection{Coupling of the interacting chain to the leads}
The model chain is connected to source and drain metallic leads. The latter are assumed non-interacting and maintained in thermal equilibrium. Absence of electron-cavity interactions in the leads is justified by  strong field screening by the metals. Effect of the leads on the conductor under study is typically described by the self-energy operator $\hat{\Sigma}= \hat{\Sigma}_{s} + \hat{\Sigma}_{d}$, where $\hat{\Sigma}_{s}$ and $\hat{\Sigma}_{d}$ govern the effects of source and drain contacts, respectively~\cite{Datta2005}. The precise form of $\hat{\Sigma}_{s}$ and $\hat{\Sigma}_{d}$ depends on the density of states in the leads and details of the lead-channel coupling. The simplifying property of these operators is they act only on the terminal sites of the chain. In other words, the matrices of $\hat{\Sigma}_{s}$ and $\hat{\Sigma}_{d}$ in the tight-binding representation have a single non-zero element each.

After the functional form of self-energy is specified, all the necessary information about transport would be encoded in propagators (Green's functions) $\hat G (E)$. They are determined from
\begin{equation}
\label{EQ-Greens-functions}
    \hat G(E) = [E \hat{I} - \hat{\mathcal H }- \hat \Sigma]^{-1},
\end{equation}
where $E$ is the energy and $\hat{I}$ is the identity operator.

From this point, our method of considering electron-boson interactions would differ from the conventional diagrammatic expansions of the electron's Green's function in powers of interaction strength~\cite{Arseev_2010,Chen_STM}. We aim at the non-perturbative account of electron-cavity couplings. Therefore, all Green's functions will be defined in the states' space spanned by electron and cavity degrees of freedom. In other words, we shall deal with new quasiparticles, the electrons dressed in $N$ cavity photons. The state of such quasiparticles will be characterized by a multi-index $\ket{\alpha} = \ket{i,N}$, where $i$ is the position in the atomic chain and $N$ is the number of photons in the cavity mode. The energy argument of the Green's function $E$ is the total energy of the dressed electron. In the absence of interactions, is comprised of purely the electronic part $\varepsilon$ and the photonic part $N\hbar\omega$. 

The extended states' space for the dressed electron can now be represented by a two-dimensional network shown in Fig.~\ref{fig1} (b). Each horizontal line corresponds to a constant number of cavity photons. Each vertical line corresponds to an electron at a given atomic site, with different possible numbers of photons in the cavity. Importantly, each $N$-th horizontal line of this network is coupled to the leads, which implies the possibility of inelastic processes, i.e. retaining the cavity in the excited state after the electron passage. The paths between source and drain across this network has a transparent physical interpretation. As example, the path starting at $\ket{N}=\ket{0}$ and ending at $\ket{M}$, $M\neq 0$ corresponds to an electron entering the cavity in the photonic ground state, and leaving the cavity with $M$ excited photons [red line in Fig.~\ref{fig1} (b), tunneling with photon emission].

Evaluation of the Green's function of the compound quantum system (\ref{EQ-Greens-functions}) requires the knowledge of the Hamiltonian and self-energy matrix elements. Both are evaluated in two steps, first with respect to cavity states, and second with respect to electron states. The elements of $\mathcal{H}$ with respect to cavity states $\bra{M}{\hat{\mathcal H}}\ket{N}$ are computed analytically with the help of Baker-Campbell-Hausdorff formula (Appendix A). In tight-binding sub-space, $\mathcal{H}$ is represented by an ordinary tridiagonal matrix.

Explication of the self-energy operator $\hat{\Sigma}$ for the interacting chain represents a more complex problem. The reason is that an electron can leave the channel either by retaining the cavity in its initial state, or by emitting/absorbing $N$ cavity photons [Fig.~\ref{fig1} (b)]. The rate at which the electron leaves the cavity should depend only on the the electron's energy $\varepsilon = E - N \hbar \omega$. The state of the cavity should not affect the possibility of electron escape. An electron is unable to leave the cavity if its energy $E-N\hbar\omega$  lies outside of the conduction band in the leads, as there's no propagating state in the leads with that energy. Assume now that the self-energy for the non-interacting chain is known and equals $\hat{\Sigma}_0(E)$ in the tight-binding representation. The self-energy for the interacting chain $\hat{\Sigma}(E)$, satisfying the above requirements, would have the following matrix element with respect to the cavity states:
\begin{equation}
\label{eq-sigma-photonic}
   \bra{M} \hat{\Sigma}(E) \ket{N} = \delta_{NM} \hat{\Sigma}_0(E-N\hbar\omega).
\end{equation}
The physical meaning of Eq. (\ref{eq-sigma-photonic}) is clarified by recalling that imaginary part of self-energy is proportional to the rate of electron escape from the studied system. Mathematically, the statement is formulated as ${\rm Im}\Sigma_0(\varepsilon) \sim \hbar v(\varepsilon)/a$, where $v(\varepsilon)$ is the electron velocity, and $a$ is the inter-atomic distance. Equation (\ref{eq-sigma-photonic}) simply represents the fact that emission of $N$ cavity photons reduces the electron velocity from $v(E)$ down to $v(E - N\hbar\omega)$, and thus reduces the rate of electron escape from the chain. If $E - N\hbar\omega < 0$, the velocity becomes imaginary, which reflects the energy constraint on the electroluminescence.

The general property of the self-energy operator without cavity coupling $\hat{\Sigma}_0$ is its action only on the terminal atoms in the chain. Formally, it can be always split into source and drain parts, such that
\begin{gather}
    \hat{\Sigma}_0 = \hat{\Sigma}_S + \hat{\Sigma}_D,\\
    \bra{i} \hat{\Sigma}_S (E)\ket{j} = \delta_{ij}\delta_{i,1} g_s(E),\\
    \bra{i} \hat{\Sigma}_S (E)\ket{j} = \delta_{ij}\delta_{i,n_{\rm at}} g_d(E).
\end{gather}
The particular form of scalar coupling functions $g_{s/d}$ on the density of states in the leads and microscopic details of atomic bonds. We adopt these in the simplest approximation of one-dimensional contact with the same bandwidth as the channel~\cite{Datta2005},
\begin{equation}
    g_s(E) = g_d(E) = \frac{2t^2}{E-2t+i\sqrt{E(4t-E)}}.
\end{equation}

\subsection{Generalized Landauer formula for the current in the interacting chain}
The current $I_0$ in a non-interacting one-dimensional system biased by voltage $eV_{sd} = \mu_s - \mu_d$ is given by the Landauer formula~\cite{Landauer}:
\begin{equation}
\label{eq-Landauer-bare}
  I_0 =\frac{2e}{h}\int\limits_{-\infty }^{+\infty }{dE\left[ n_F\left( E \right)-n_F\left( E-e{{V}_{sd}} \right) \right]{\mathcal T}_{0}\left( E \right)}, 
\end{equation}
where $n_F(E)$ are the fermionic occupation numbers. The energy-dependent transmission probability in the non-interacting chain ${\mathcal T}_{0} \left( E \right)$ can be expressed via~\cite{Caroli}
\begin{equation}
\label{eq-Caroli}
   {\mathcal T}_{0} ={\rm Tr} [\hat\Gamma_S \hat G \hat \Gamma_D \hat G^+ ]
\end{equation}
where $\hat{G}$ is the Green's function in the tight-binding representation, $^+$ stands for Hermitian conjugate, and $\hat\Gamma_{s/d}$ are the rate matrices of electron exchange with source and drain,
$\hat\Gamma_{s/d} = i [\hat{\Sigma}_{s/d} - \hat\Sigma^+_{s/d}]$~\cite{Datta2005}. The expression for transmission ${\mathcal T}_{0}$ is greatly simplified in 1d atomic chains, where $\hat{\Sigma}$-operators act only at the terminal sites~\cite{Caroli}
\begin{equation}
\label{eq-Caroli-simple}
    {\mathcal T}_{0}(E) = 4 {\Im} g_s(E) {\Im} g_d(E) |\bra{1}\hat{G}(E)\ket{n_{\rm at}}|^2;
\end{equation}
here $\Im$ stands for the imaginary part. In other words, transmittance is proportional to the modulus squared of source-drain propagator $|\bra{1}\hat{G}\ket{n_{\rm at}}|^2$.

Our next goal is to modify the Eqs.~(\ref{eq-Landauer-bare}) and (\ref{eq-Caroli}) accounting for electron-cavity interactions to an arbitrary order of the interaction strength. The first necessary building block here is the density matrix of the system ''electron in the leads + cavity''. We assume that the electromagnetic field does not penetrate into the metal leads, while strong electron scattering quickly destroys the electron-photon coherence. It implies that electronic and photonic degrees of freedom are decoupled in the leads and, moreover, obey the Gibbs distribution. Formally, the elements of density matrix in the leads between the states $\ket{\alpha} =\ket{i,N} $ and $\ket{\alpha'} = \ket{i',N'}$ are given by 
\begin{equation}
    \bra{\alpha'}{\hat{\rho }}\ket{\alpha}= \bra{i'} {\hat{\rho }}_{e}\ket{i} \otimes  \bra{N'} \hat{\rho}_{ph}\ket{N}.
\end{equation}
The 'photonic part' of the density matrix is given by the ordinary Gibbs distribution for $N$ photons in the mode
 \begin{equation}
      \bra{N'} \hat{\rho}_{ph}\ket{N} = \frac{1}{Z}{{\delta }_{N{N}'}}\exp \left( - N\hbar \omega /kT \right),
 \end{equation}
$Z=[1-{{e}^{-\beta \hbar \omega }}]^{-1}\equiv n_B(\omega ) + 1 $ is the statistical sum for an individual cavity mode, which is the Bose function of energy $\hbar \omega $. The electron density matrix obeys the Fermi distribution with energy $\varepsilon $ and chemical potential $\mu $:
\begin{equation}
    \bra{i'} {\hat{\rho }}_{e}\ket{i} =\int\limits_{-\infty }^{+\infty }{d\varepsilon {{A}_{ii'}( \varepsilon )n_F(\varepsilon)}},
\end{equation}
here ${A}_{ii'}\left( \varepsilon  \right)$ is the electron’s spectral function in the leads, and $n_F(\varepsilon) = [e^{(\varepsilon -\mu )/kT } +1]^{-1}$ is the electron's Fermi function.

Absence of electron-cavity interactions in the leads enables us to introduce the well-defined occupation numbers for quantum states with given total energy $E$ and photon number $N$:
\begin{equation}
n(E,N) = n_F(E - N \hbar\omega) \times \frac{e^{-N\hbar\omega/kT}}{n_B(\omega)+1}.
\end{equation}
Their meaning is the probability of finding an electron with energy $\varepsilon_e = E - N \hbar\omega$ dressed into $N$ cavity photons. Summation of $n(E,N)$ with respect to photon numbers $N$ yields $n_F(\varepsilon_e)$, which reflects the single-electron nature of the problem. The Landauer-like formula can now be written down for new non-interacting excitations, the electrons dressed into multiple cavity photons, as (see Appendix B):
\begin{equation}
\label{eq-Landauer-gen-Fermi}
I  =  \frac{2e}{h} \sum_{N, M} \int d E [n_s (E , N) - n_d (E , M)]
  {\mathcal{T}}_{N M} (E), 
\end{equation}
where $n_{s/d}$ are the occupation numbers for the dressed electrons in the source and drain. In the Boltzmann limit, the occupation numbers of dressed electrons $n(E , N)$ would depend only on total energy, $n(E,N) \approx e^{-(E-\mu)/kT}/[n_B(\omega) + 1]$. The generalized Landauer formula becomes especially simple
\begin{equation}
\label{eq-Landauer-like}
    I=\frac{2e/h}{n_B(\omega ) + 1}
    \int\limits_{-\infty }^{+\infty }{dE\left[ f_s\left( E \right)-f_d\left( E \right) \right]\sum\limits_{N,M=0}^{+\infty }{\mathcal{T}}_{NM}( E )},
\end{equation}
where $f_{s/d}(E) = e^{-(E-\mu_{s/d})/kT}$ is the Boltzmann exponent for the energy $E$. Above, ${\mathcal{T}}_{NM}( E )$ is the probability of electron transmission from source with $N$ photons in the cavity to the drain with $M$ photons in the cavity. The expression for photon-number-resolved transmittance generalizes the results (\ref{eq-Caroli}-\ref{eq-Caroli-simple}):
\begin{multline}
\label{eq-Caroli-photons}
\mathcal{T}_{NM} =\\
{\rm Tr} [ \bra{N}{\hat\Gamma}_S\ket{N} \bra{N} \hat G \ket{M} \bra{M} \hat \Gamma_D\ket{M} \bra{M}\hat G^+\ket{N} ] = \\
4 {\Im} g_s(E-N\hbar\omega) {\Im} g_d(E - M\hbar\omega) |\bra{N,1}\hat{G}\ket{n_{\rm at},M}|^2
\end{multline}
The diagonal elements $\mathcal{T}_{NN}$ show the probabilities of electron transfer between source and drain without real photon excitation. Yet, virtual photon emission along the path is fully included into $\mathcal{T}_{NN}$. The off-diagonal elements $\mathcal{T}_{NM}$ stand for probabilities of assisted tunneling, $M-N$ is the number of emitted/absorbed real photons during a single electron passage. 

The modified Landauer formula (\ref{eq-Landauer-like}) is applicable only at low fermion occupation numbers $n_F \ll 1$, otherwise, the Pauli blocking principle would hinder the emission of cavity photons~\cite{Ciuti_2022}. Formally, this corresponds to the Fermi levels $\mu_s$ and $\mu_d$ lying outside of the conduction band. The current in such situation is carried by thermally excited carriers with $E > \{ \mu_s, \mu_d\}$.

The practically measured quantity is the conductance $G = \partial I/\partial V_{SD}$. To single out the contributions of photon-assisted and elastic transport, we endow $G$ with photon indices $G_{NM}$ and define it according to:
\begin{equation}
\label{eq-conductance-nm}
    G_{NM} = \frac{G_Q}{n_B(\omega) + 1}  \int{\mathcal{T}_{NM}(E)e^{\frac{E}{kT}} \frac{dE}{kT}}.
\end{equation}
The physical meaning of $G_{NM}$ is the part of total conduction, where an electron enters the cavity with $N$ photons, and emits/absorbs $M-N$ photons during its path. Above, we have introduced $G_Q = [2 e^2/ h] e^{-\mu/kT}$, the conductance quantum timed by the Boltzmann distribution at the band edge, $\mu = (\mu_s + \mu_d)/2$. The Fermi level $\mu$ in the non-degenerate semiconductor affects the conductivity only via a multiplicative factor $e^{-\mu/kT} < 1$.

\begin{figure*}[ht] 
    \includegraphics[width=1.0\linewidth]{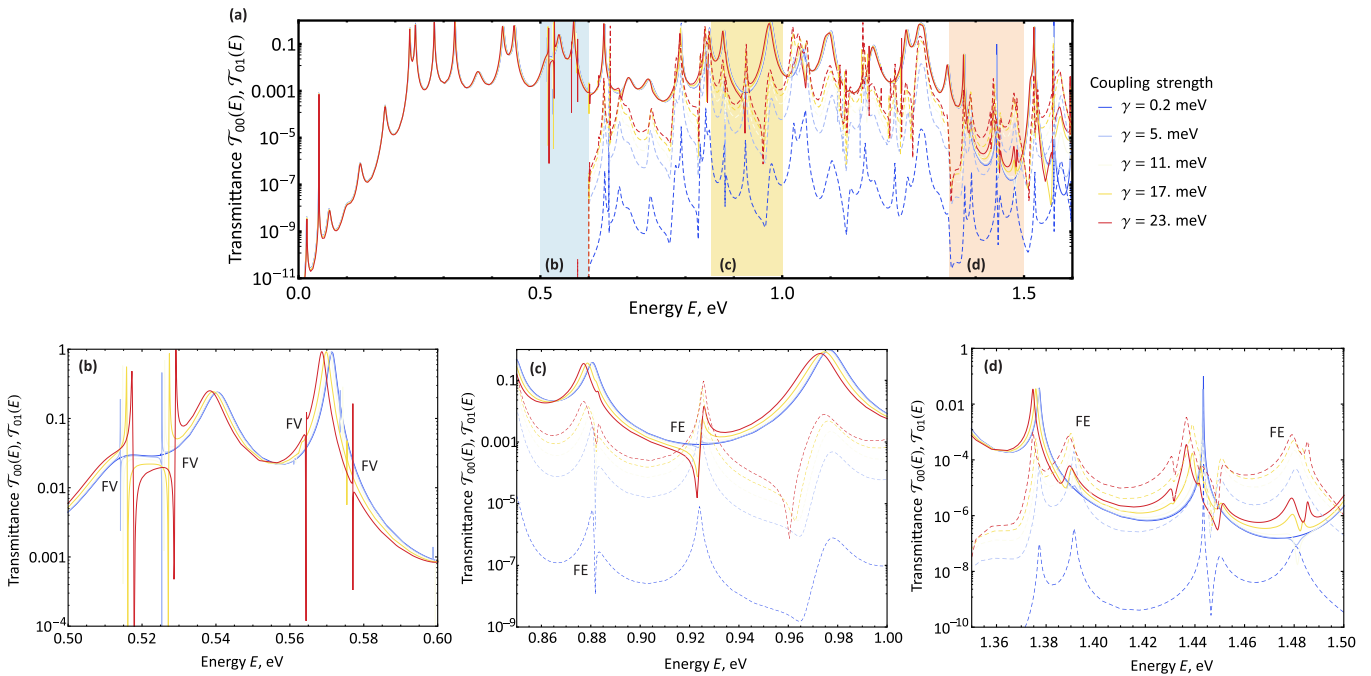}
    \caption{{\bf Effect of cavity on the electron transmission through a disordered 1d chain.} (a) Energy-dependent transmission coefficients without real photon emission $\mathcal{T}_{00}$ (solid) and with single real photon emission $\mathcal{T}_{01}$ (dashed). The coupling strength varies from $0.2$ meV to $23$ meV, photon energy is $\hbar\omega = 0.6$ eV, disorder amplitude is $\delta U = 0.2$ eV. (b-d) Magnified views of transmittance in the selected energy sectors. (b) In the low-energy region $E<\hbar\omega$, an electron can emit a virtual photon and travel along a bound state in the disorder potential. This results in Fano interference of original path and path with virtual photon emission (denoted as FV). (c) At intermediate energies, $E \gtrsim \hbar\omega$, emission of cavity photon pushes the electron to a quasi-bound state weakly coupled to the leads. This results in a broad Fano resonance with appearance of electroluminescence (denoted as FE). (d) At high energies, $E\sim 4t$, an emission of cavity photon pushes the electron into the state decaying rapidly to the leads. This results in dominance of electroluminescence. }
    \label{fig-high-photon}
\end{figure*}

\section{Results: Modifications of one-dimensional conductance due to cavity coupling}
\label{sec-Results}

We proceed to study the effects of cavity coupling on energy-resolved transmittance $\mathcal{T}(E)$ and thermally-averaged conductance $G$ in one-dimensional atomic chains. These quantities were obtained via a numerical computation of Green's functions \ref{EQ-Greens-functions} and subsequent evaluation of transmittance (\ref{eq-Caroli-photons}) and conductance (\ref{eq-conductance-nm}). The number of excited cavity states is chosen adaptively to ensure the convergence of transmission coefficient $\mathcal T_{00}$ with 1 \% accuracy. In the following calculations, such convergence required from 1 to 9 excited cavity states, depending on the coupling strength $\gamma$.

We start the discussion of our numerical results with consideration of relatively high photon energy $\hbar\omega = 0.6$ eV, which is comparable to the bandwidth $4t=1.6$ eV and much exceeds the disorder amplitude $\delta U = 0.2$ eV. The chain length is taken to be $n_{\rm at} = 60$. The computed plots of transmittances $\mathcal T_{00}(E)$ (ground-state transmittance) and $\mathcal T_{01}(E)$ (transmittance with one-photon emission) is shown in Fig.~\ref{fig-high-photon} (a). Sub-panels (b)-(d) display the magnified views of transmittance in several characteristic energy ranges. In the absence of cavity, the transmittance $\mathcal T_{00}(E)$ displays several sharp resonances corresponding to the electron passage along the quasi-bound states $E_i$ in the random potential. The transmittance envelope drops to zero at both lower and upper band edges due to predominant localization of states with small velocity.

The main effects of cavity on transport consists in (1) emergence of extra resonant structures in transmittance $\mathcal T_{00}(E)$ (2) onset of electroluminescence $\mathcal T_{01}(E)$ at energies $E > \hbar\omega$. Each new resonant structure appears approximately at the energy $E = E_i + \hbar\omega$. In other words, it represents a photon replica of an original resonant path. The two electron paths -- one without photon emission and the other with photon emission and re-absorption -- interfere with each other. This results in emergence of a characteristic Fano structure in an original transmission curve $\mathcal T_{00}(E)$. The width and amplitude of the new Fano resonance depends on the position of the electron state $E$ with respect to the mid-band.

\begin{figure*}[ht] 
    \includegraphics[width=1.0\linewidth]{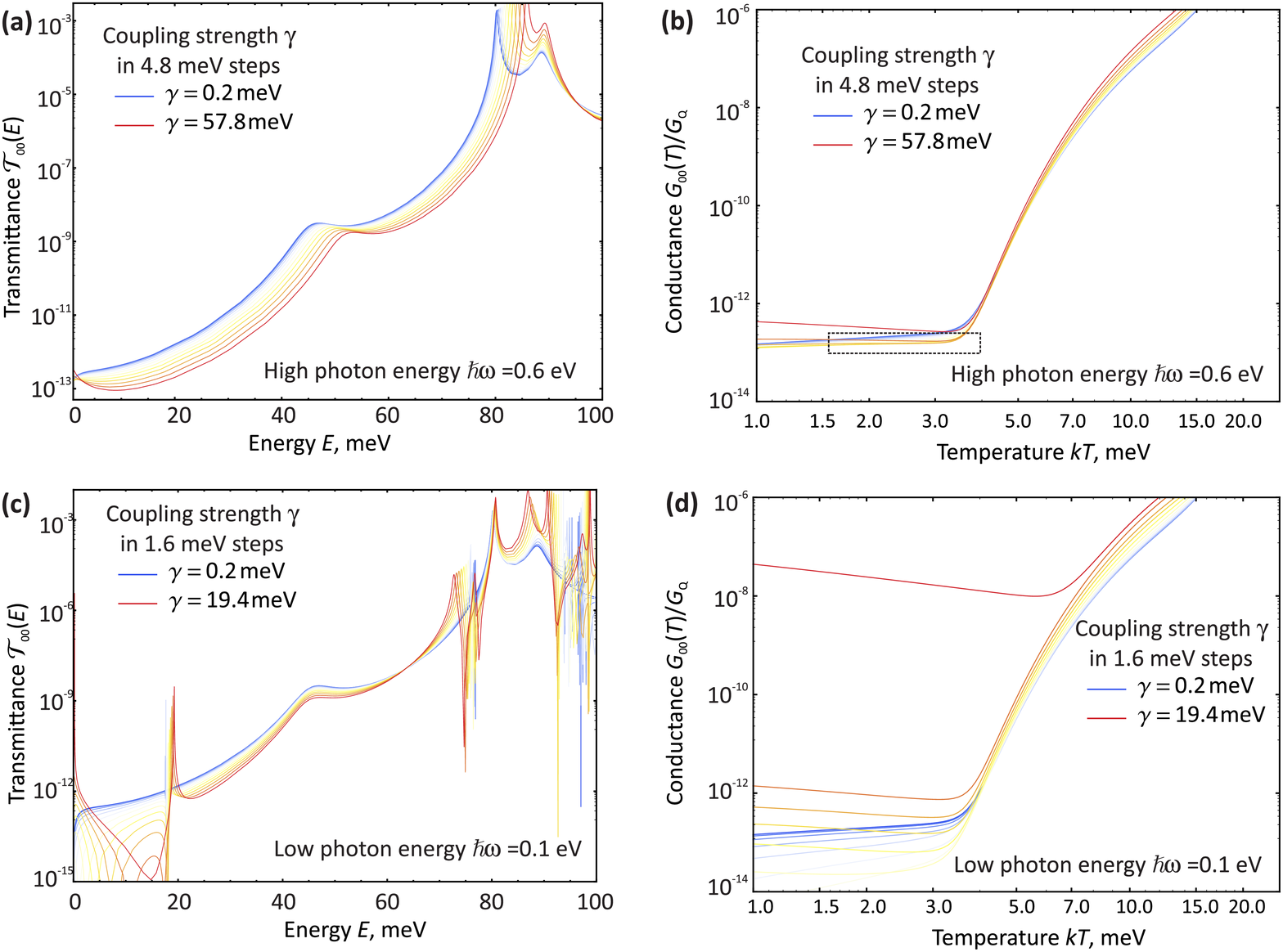}
    \caption{{\bf Cavity-induced changes in energy-resolved transmittance and in thermally averaged conductance.} The resuts are obtained for an atomic chain with $n_{\rm at}=60$, disorder amplitude $\delta U = 0.2$ eV, bandwidth $4t=1.6$ eV. (a,b) Simulation for photon energy $\hbar\omega = 0.6$ eV, which is larger than the disorder amplitude. (c,d) The same disorder realization simulated for photon energy $\hbar\omega = 0.1$ eV, smaller than the disorder amplitude. Left and right columns show the energy-resolved transmittance and the respective thermally-averaged conductance.}
    \label{fig-low-vs-high}
\end{figure*}

In the high-energy sector, $E \lesssim 4t$, the Fano structures in the transmittance $\mathcal{T}_{00}$ are broad and faint. At the same time, the amplitude of electroluminescence $\mathcal{T}_{01}$ is relatively high and may exceed the zero-photon transmission at relatively small coupling [$\gamma^* \approx 10$ meV in Fig.~\ref{fig-high-photon} (d)]. The zero-photon path $\mathcal{T}_{00}$ in this situation is suppressed due to the carrier localization at the upper mobility edge, and the original bound states are weakly coupled to the leads. After cavity photon emission, an original electron gets closer to the mid-band, where the group velocity and coupling to the leads are large. These final states have relatively short lifetimes and decay quickly by releasing the electron to the leads. In such a situation, the electroluminescence process readily dominates the zero-photon transmission.

At intermediate energies, $E\sim 2t$, an original electron has relatively large group velocity and is less prone to localization in a disorder potential. Emission of a virtual photon now pushes the electrons closer to the band bottom, i.e. to the states weakly coupled to the leads. The weak coupling of the final state to the leads results in a narrow Fano resonance in the original transmission curve  $\mathcal{T}_{00}(E)$ [Fig.~\ref{fig-high-photon} (c)]. The coupling of the final state to the leads also results in electroluminescence peak, albeit weaker than in high-energy sector.  

At even lower energies $E<\hbar\omega$, an electron is incapable of real photon emission. Nevertheless, its transmission curve $\mathcal{T}_{00}(E)$ acquires extra Fano resonant structures due to emission of virtual photons [FV-peaks in Fig.~\ref{fig-high-photon} (b)]. The very possibility of these low-energy peaks is enabled by the disordered character of the potential. A disordered {\it non-interacting} chain has several energy levels truly bound in the random potential with energies $E_i < 0$. An electron incident on the {\it interacting chain} can gain access to these levels via emission and re-absorption of the virtual cavity photons [process \# 1 in Fig.~\ref{fig1} (a)]. Finite linewidth of these Fano-type structures $\sim \gamma^2/\omega$ is assured by light-matter coupling itself, and is thus very small.

The studies of cavity effects on electron transmission $\mathcal{T}(E)$ in a broad energy range represent merely an academic interest. Indeed, all electron states relevant to transport in non-degenerate semiconductors are located at energies $E \sim kT$ from the lower band edge. Therefore, only the modifications of the low-energy transmittance $\mathcal{T}(E)$ can change the thermally-averaged conductance $G(T)$.

To study the modifications of transmittance relevant to the conductivity of non-degenerate semiconductors, we focus at the low-energy sector of the simulated $\mathcal{T}_{00}(E)$. We first retain the same photon energy $\hbar\omega = 0.6$ eV and plot the results in panels (a-b) of Fig.~\ref{fig-low-vs-high}. The first effect of cavity in this sector is represented by a slight {\it upward} spectral shift of the transmission resonances. The sign of the shift may look unusual because second-order perturbative correction 
\begin{equation}
\label{eq-shift-1}
    \delta^{(1)} E_i = \frac{e^2}{m^2c^2}\sum_{k}{\frac{ |p_{ik}|^2 A_{\rm vac}^2}{E_i - (E_k + \hbar\omega)}}
\end{equation}
to the energy of the ground state should be always negative. The explanation of the positive shift lies in the the two facts. First, the lowest resonance in  $\mathcal{T}_{00}(E)$ does not necessarily correspond to the lowest energy level of the interacting chain. As discussed above, even lower levels bound in the random potential are present at $E<0$. Interaction of the lowest resonance with these levels may lead to the positive energy shifts. 

Second, apart from 'virtual photon' correction to the energies given by (\ref{eq-shift-1}), there exists another 'diamagnetic' correction. It is also quadratic in the electron-photon coupling and appears in the first order of the perturbation theory:
\begin{equation}
\label{eq-shift-2}
    \delta^{(2)} E_i = \frac{e^2}{2mc^2} A_{\rm vac}^2 = 2\frac{\gamma^2}{t}.
\end{equation}
As a result of the two above-mentioned perturbations, most resonances in $\mathcal{T} (E)$ in the low-energy sector for disordered chains are generally shifted upwards in energy with increasing the cavity coupling $\gamma$. This results in reduced thermally-averaged transmission $\langle \mathcal{T}\rangle_{th}$  at low temperatures with increasing the coupling strength $\gamma$. This range of temperatures and coupling strengths is shown in Fig.~\ref{fig-low-vs-high} (b) with a dashed rectangle.

The second effect of cavity coupling is also related to positive energy shifts of levels with increasing the coupling strength $\gamma$. It lies in the elevation of energy levels from the band of strictly localized states $E<0$ to the band of accessible energies $E>0$. More precisely, a discrete level in the disorder potential with a low binding energy $-|E_b|$ may acquire a positive correction (\ref{eq-shift-2}). Its net final energy $-|E_b| + \delta ^{(1)} E + \delta^{(2)} E > 0$ thus becomes accessible to the carriers incident from the leads. Such a 'prolumination' of the energy levels by cavity results in emergence of a resonance in transmission $\mathcal{T}_{00}(E)$ at nearly-zero energy [see the red curve in Fig.~\ref{fig-low-vs-high} (a) in the vicinity of $E\approx 0$]. The low-energy resonance, in turn, implies a positive correction to the conductance at low temperatures, $\left. \partial G/\partial \gamma \right|_{T \rightarrow 0} >0$. While the low-temperature conductance is enhanced via coupling to the cavity, the temperature dependence of conductance becomes anomalously decaying, $\partial G/\partial T <0$, see the negative slope of the red curve in Fig.~\ref{fig-low-vs-high} (b). The latter effect stems from an abrupt decrease in $\mathcal{T}_{00}(E)$ with energy at the right shoulder of the transmission resonance. 

\begin{figure*}[ht] 
    \includegraphics[width=1.0\linewidth]{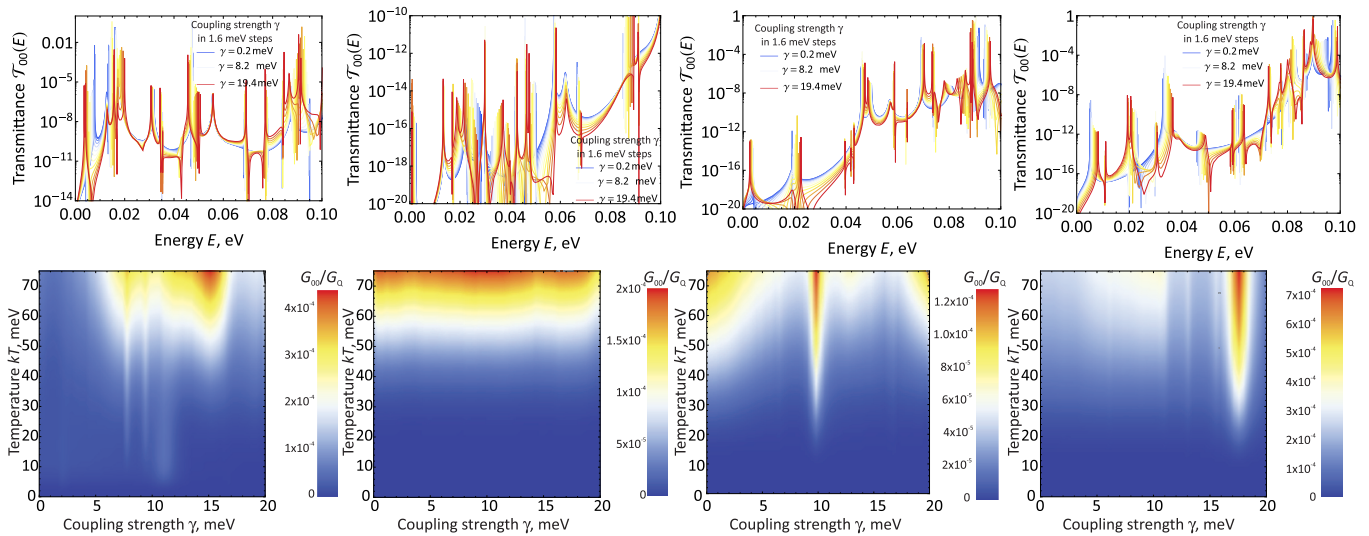}
    \caption{{\bf Effect of cavity on 1d localization in long conductors.} Simulated energy-resolved transmittance $\mathcal{T}_{00}(E)$ (top row) and thermally-averaged conductance $G_{00}(T,\gamma)$ for various realizations of disorder. Chain length $n_{\rm at} = 120$, which is two times longer than that in Figs.~\ref{fig-high-photon} and \ref{fig-low-vs-high}. The conductance generally reaches a maximum at some $\gamma$ corresponding to the passage of bound state through the conduction band bottom. Photon energy $\hbar\omega=0.1$ eV, disorder amplitude $\delta U=0.2$ eV}
    \label{fig-long-chain}
\end{figure*}

The modification of conductance associated with 'prolumination' of the weakly bound states is manifested more sharply with decreasing the cavity photon energy. This is best illustrated in Fig.~\ref{fig-low-vs-high} (c-d), where the same realization of disorder is studied at cavity photon energy $\hbar\omega=0.1$. The zero-energy resonance in transmission becomes visible already at coupling strength $\gamma^*\sim 10$ meV, which is six times smaller than the corresponding $\gamma^*$ at high photon energy $\hbar\omega=0.6$ eV. The reason for stronger cavity effects at low photon energies lies in smaller energy denominators in (\ref{eq-shift-1}), and thus in larger perturbative energy shifts at a fixed value of $\gamma$~\footnote{Increased strength of cavity effects at $\hbar\omega \rightarrow 0$ persists if only $\gamma$ itself is independent of frequency. In fact, $\gamma \propto A_{\rm vac} \propto (\omega V)^{-1/2}$. Low-frequency cavities have larger size $V$, and in fact $\gamma \propto \omega$.}. Further increase in $\gamma$ shifts the resonance away from $E \approx 0$, and therefore reduces the conduction.

Another set of modifications in $\mathcal{T}_{00}(E)$ at the band bottom is associated with emergence of sharp Fano resonances. An electron at nearly-zero energy $E \approx 0$ is now capable of emitting the virtual photon and pass to the discrete level in the disorder potential. The sharp Fano structure discussed above can now appear right at the lower band edge, as illustrated in Fig.~\ref{fig-low-vs-high} (c). Though the Fano structures modify the transmission spectra, their effect on the $T$-dependent conductance is non-univocal. Indeed, a sharp increase in $\mathcal{T}_{00}(E)$ in a narrow range of energies can be compensated by a transmission drop at the other energies. Further examples of cavity modifications of conductance for various disorder realizations can be found in the Supplementary material~\cite{Suppinfo}.

Most of the cavity effects on conductance seen in the vicinity of the band bottom are associated with the presence of bound states in the disordered potential. It is natural to assume that an increase in the chain length, at the same disorder statistics, would enhance the number of bound levels. As soon as the motion of electron remains coherent throughout the chain, longer chains should be more susceptible to cavity effects. This suggestion is fully confirmed in Fig.~\ref{fig-long-chain}, where we present the transmission simulations for a chain of length $n_{\rm at} = 120$, twice longer than that in Figs.~\ref{fig-high-photon}-\ref{fig-low-vs-high}, and at four different realizations of disorder. The number of transmission resonances (both ordinary and Fano-type) per unit energy window is here much bigger than that for short chains. For most disorder realizations, a weakly-bound state crosses the band bottom $E=0$ as the coupling constant increases. This effect results in non-monotonic $\gamma$-dependence of the thermally-averaged conductance.

\section{Discussion of the results}
\label{sec-discussion}
We have revealed several effects of electron-photon coupling leading to the modification of conductivity in the localized regime for non-degenerate carrier statistics. The first class of modifications is associated with Fano resonances in transmission associated with the emission of virtual cavity photons and passage of electron along the bound levels in the disorder potential. The effect is observed only at small photon energies compared to the disorder amplitude $\hbar\omega \lesssim \delta U$. An extra requirement for this effect is the absence of degeneracy throughout the bound states, $\mu < \delta U$. Otherwise, Pauli blocking poses a strong restriction on the photon emission processes, be they real or virtual. 

The second class of cavity effects on the one-dimensional conductivity lies in potentially strong shifts of bound and quasi-bound energy levels with increasing the coupling strength. The cavity-induced diamagnetic correction to the energy levels is always positive, while the virtual-photon correction has an indeterminate sign. The cavity-induced energy shifts are most pronounced in conduction if an initially bound state becomes quasi-bound and accessible to electrons incident from the leads. Such energy shift implies the cavity-enhanced conduction at low temperatures. Observation of this effect requires sharp conduction band edge in the leads -- otherwise, no well-pronounced boundary between bound and quasi-bound states would exist. 

While the present discussion was limited to the one-dimensional systems, we can suggest that the above cavity effects are generic and apply also to the two- and three-dimensional systems. The effect of cavity on the energy levels in 2d systems should be especially strong, as the binding energy in the disordered 2d system is exponentially small. We may thus suggest that placement of disordered 2d systems into cavities should delocalize the weakly bound states and make them contribute to transport.

Finally, we estimate whether the obtained values of coupling constant $\gamma \sim 0.1$ meV necessary to observe the cavity effects on localized conductance are realistic. Relating $\gamma$ to cavity parameters, and assuming the cavity to have the volume $V$, we get
\begin{equation}
\label{eq-gamma-numerical}
    \gamma = \left({\frac{e^2}{\hbar c}}\right)^{1/2} \frac{\hbar^2}{2m^* a}\left(\frac{\lambda}{V}\right)^{1/2},
\end{equation}
where we have introduced the effective mass according to $t=\hbar^2/2m^*a^2$. Taking the minimum possible mode volume $V=\lambda^3$, we simplify (\ref{eq-gamma-numerical}) to
\begin{equation}
    \gamma = \left({\frac{e^2}{\hbar c}}\right)^{1/2} \frac{\hbar^2}{2m^* a \lambda}.
\end{equation}
The numerical estimate for $m^* = 0.067 m_0$ (GaAs quantum wire), $a=0.5$ nm and $\lambda = 10$ $\mu$m provides $\gamma = 0.1$ meV. Fortunately, this corresponds to some pronounced effects of cavity on conduction at the lower mobility edge. Higher values of $\gamma$ may be obtained in 2d plasmonic cavities~\cite{Singe_GR_cavities}. The characteristic wavelength of 2d plasmons is $\sim 10^2$ times below the free-space photon wavelength. With this 'compression ratio', 2d plasmonic cavities can provide $\sim 10^3$-fold enhancement of the coupling constant. Another approach toward increasing the coupling strength lies in the placement of 1d conductors into the cavities with field singularities. The most prominent example of such cavity is represented by a slot in the planar metallic pad~\cite{Faist_magneto_transport,Faist_QH_breakdown}. Though the average amplitude of fluctuations in such cavity is still given by $A_{\rm vac} = (2\pi \hbar c/\omega V)^{1/2}$, the local values of field near the slot edges can take on much higher values.

The computed temperature-dependent conduction of 1d chains in a cavity $G(T,\gamma)$ can be numerically very different from the respective quantity in the uncoupled structure, $G(T,\gamma = 0)$. However, the temperature dependence $G(T)$ in an interacting chain does not possess any distinctive features that were absent at $\gamma = 0$. To reveal the effect of coupling, one has to perform conduction measurements at variable coupling strength. Fortunately, the corresponding setups based on movable Bragg mirrors are available~\cite{Bragg_movable_1,Bragg_movable_2}.

The work of DS and GA was supported by the grant 22-29-01034 of the Russian Science Foundation. LMM acknowledges Project PID2020-115221GB-C41, financed by MCIN/AEI/10.13039/501100011033, and the Aragon Government through Project Q-MAD. The authors thank V.Vyurkov and V.Muravev for fruitful discussions.

\bibliography{Sample}

\begin{widetext}
\appendix
\section{Numerical realization of electron-cavity Hamiltonian}
\label{app-Hamitonian}

For computational purposes, it would be necessary to obtain the representation of the Hamiltonians in the matrix form. Below, we establish the structure of the respective matrices. In the interacting channel, the photon and electron degrees of freedom are coupled. The wave function of the compound system 'electron + field' (if its state can be described with the wave function) can be presented as
\begin{equation}
   \Psi \left( x_{i},N \right)=\sum\limits_{N=0}^{\infty }{{c_N}\left( {{x}_{i}} \right)\left| N \right\rangle }. 
\end{equation}
In the exact diagonaization approach, we will limit our consideration to several lowest excited states of photon mode, so that
\begin{equation}
\Psi \left( x_{i}, N \right)={{c}_{0}}\left( {{x}_{i}} \right)\left| 0 \right\rangle +{{c}_{1}}\left( {{x}_{i}} \right)\left| 1 \right\rangle +{{c}_{2}}\left( {{x}_{i}} \right)\left| 2 \right\rangle +...    
\end{equation}
After such truncation, electron-photon states in the channel are described by a vector of length ${{N}_{ph}}\times n_a$, where $n_a$ is the number of atomic sites and ${{N}_{ph}}$ is the number of photonic states. The size of corresponding Hamiltonian matrix is $\left( {{N}_{ph}}\times n_a \right)\times \left( {{N}_{ph}}\times n_a \right)$. In all our numerical calculations, we arrange the Hamiltonian matrices in the block form
\begin{equation}
    {{\hat{\mathcal H}}}=\left( \begin{matrix}
   \left\langle  0 \right|{\hat{ \mathcal H}}\left| 0 \right\rangle  & \left\langle  0 \right|{\hat{ \mathcal H}}\left| 1 \right\rangle  & \left\langle  0 \right|{{{\hat{\mathcal H}}}}\left| 2 \right\rangle  & ...  \\
   \left\langle  1 \right|{\hat{\mathcal H}}\left| 0 \right\rangle  & \left\langle  1 \right|{{{\hat{\mathcal H}}}}\left| 1 \right\rangle  & \left\langle  1 \right|{{{\hat{ \mathcal H}}}}\left| 2 \right\rangle  & ...  \\
   \left\langle  2 \right|{\hat{\mathcal H}}\left| 0 \right\rangle  & \left\langle  2 \right|{{{\hat{\mathcal H}}}}\left| 1 \right\rangle  & \left\langle  2 \right|{{{\hat{\mathcal H}}}}\left| 2 \right\rangle  & ...  \\
   ... & ... & ... & ...  \\
\end{matrix} \right),
\end{equation}
where brakets $\bra{N}{\hat{\mathcal H}}\ket{N'}$ stand for taking the matrix elements between 
$N$-th and $N'$-th states of electromagnetic field. Each element of the block matrix above is the matrix of size $n_a \times n_a $, it acts already in the tight-binding space. 

Below, we present the block representation of operators constituting $\hat {\mathcal H}$. The field term is diagonal,
\begin{equation}
\bra{N} {\hat{\mathcal H}}_f \ket{N'} = \hat{I} \delta_{NN'} N  \hbar\omega,
\end{equation}
where the identity operator $\hat{I}$ acts in the tight-binding space.

The diagonal and off-diagonal matrix elements of chain Hamiltonian are different. The diagonal elements are simply
\begin{equation}
\bra{N} {\hat{\mathcal H}}_{\rm ch} (\gamma) \ket{N} = \left( \begin{matrix}
   {U_{1}} & 0 & 0 & 0 & ...  \\
   0 & {U_{2}} & 0 & 0 & ...  \\
   0 & 0 & {U_{3}} & 0 & ...  \\
   ... & ... & ... & ... & ...  \\
   ... & 0 & 0 & 0 & {U_{n_a}}  \\
\end{matrix} \right).
\end{equation}
In other words, they represent the part of tight-binding Hamiltonian responsible for on-site energies. The off-diagonal matrix elements ($N \neq N'$) are more complex
\begin{equation}
\bra{N} {\hat{\mathcal H}}_{\rm ch} (\gamma) \ket{N'} = t \left( \begin{matrix}
   0 & h_{NN'}\left(\frac{\gamma}{t}\right) & 0 & 0 & ...  \\
   h_{NN'}\left(-\frac{\gamma}{t}\right)& 0 & h_{NN'}\left(\frac{\gamma}{t}\right) & 0 & ...  \\
   0 & h_{NN'}\left(-\frac{\gamma}{t}\right) & 0 & h_{NN'}\left(\frac{\gamma}{t}\right) & ...  \\
   ... & ... & ... & ... & ...  \\
   ... & 0 & 0 & h_{NN'}\left(-\frac{\gamma}{t}\right) & 0  \\
\end{matrix} \right).
\end{equation}
Above, we have introduced an auxiliary function 
\begin{equation}
    {h_{NM}}\left( g \right)= \bra{N} \exp \left( -ig\left[ \hat{b}+{{{\hat{b}}}^{+}} \right] \right) \ket{M},
\end{equation}
which is the matrix elements of harmonic exponent between $N$-th and $M$-th states of the oscillator, $g = \gamma/t$. Explicit calculation of $h_{NM}$-factors is done with Baker-Campbell-Hausdorff formula:
\begin{equation}
    \exp\left( -ig\left[ \hat{b}+{\hat{b}^{+}} \right] \right) = e^{-g^2/2} \exp\left( -ig {\hat{b}^{+}} \right)  \exp\left( -ig {\hat{b}} \right).
\end{equation}
The exponent of annihilation operator $\exp\left( -ig {\hat{b}} \right)\ket{M}$ produces only finite sums. They can be evaluated in the closed form by expanding the exponent in Taylor series. This results in:
\begin{gather}
  {{h}_{NM}}\left( g \right)=\left\langle n|\exp \left( -ig\left[ \hat{b}+{{{\hat{b}}}^{+}} \right] \right)|m \right\rangle ={{e}^{-g^2/2}}
  \sum\limits_{s=0}^{N}{\sum\limits_{j=0}^{N}{\frac{{{\left( ig \right)}^{s}}}{s!}\frac{{{\left( ig \right)}^{j}}}{j!}{{\delta }_{N-s,M-j}}P\left( j, M \right)P\left( s, N \right)}},\quad  \\ 
 P\left( j, M \right)=\sqrt{\left[ M-\left( j-1 \right) \right]\left[ M-\left( j-2 \right) \right]...\left[ M-1 \right]M}. 
\end{gather}
For example, some lowest elements are:
\begin{gather}
    {{h}_{00}}\left( g \right)={{e}^{-{{g}^{2}}/2}},\quad {{h}_{11}}\left( g \right)={{e}^{-{{g}^{2}}/2}}\left( 1-{{g}^{2}} \right),\quad {{h}_{01}}\left( g \right)=ig{{e}^{-{{g}^{2}}/2}},\quad {{h}_{02}}\left( g \right)=-\frac{1}{\sqrt{2}}{{g}^{2}}{{e}^{-{{g}^{2}}/2}}.
\end{gather}

\section{Photon-dressed electrons and the generalized Landauer formula}

The usual Landauer's formula is derived for noninteracting electrons, therefore, we need to cast the Hamiltonian~(\ref{eq-Hamiltonian}) into a noninteracting form. We introduce auxiliary creation/annihilation operators $\hat{c}_{i N } $, $\hat{c}_{i N }^+$ describing a fictitious fermion on a 2d lattice. Electron and photon creation/annihilation operators can be written as
\begin{gather}
  \hat{a}_{i }^+ \hat{a}_j  =  \sum_{N = 0}^{\infty} \hat{c}_{i N }^+
  \hat{c}_{j N},\\
  \hat{b} =  \sum_{i = 1}^{n_A} \sum_{N = 0}^{\infty} \sqrt{N + 1}
  \hat{c}_{i N }^+ \hat{c}_{i, N + 1} 
\end{gather}

Since only one $| i N \rangle$ state can be occupied, quartic terms in the
Hamiltonian will reduce to the quadratic ones:
\begin{equation}
  \hat{c}_{1 }^+ \hat{c}_2 \hat{c}_{3 }^+ \hat{c}_4  =  \hat{c}_{1 }^+
  \hat{c}_4 \delta_{23} - \hat{c}_{1 }^+ \hat{c}_{3 }^+ \hat{c}_2 \hat{c}_4 =
  \hat{c}_{1 }^+ \hat{c}_4 \delta_{23} . 
\end{equation}
Therefore,
\begin{gather}
  \left( \sum_{j = 1}^{n_A} \sum_{N = 0}^{\infty} \sqrt{N + 1} \hat{c}_{j N
  }^+ \hat{c}_{j, N + 1} \right)^k  = \sum_{j = 1}^{n_A} \sum_{N =
  0}^{\infty} \sqrt{\frac{(N + k) !}{N!}} \hat{c}_{j N }^+ \hat{c}_{j, N + k},\\
  \exp \left( \alpha \sum_{j = 1}^{n_A} \sum_{N = 0}^{\infty} \sqrt{N + 1}
  \hat{c}_{j N }^+ \hat{c}_{j, N + 1} \right)  = \sum_{k = 0}^{\infty}
  \sum_{j = 1}^{n_A} \sum_{N = 0}^{\infty} \frac{\alpha^k}{k!} \sqrt{\frac{(N
  + k) !}{N!}} \hat{c}_{j N }^+ \hat{c}_{j, N + k}, 
\end{gather}

\begin{multline}
  e^{\pm \frac{i \gamma}{t} (\hat{b} + \hat{b}^+)}  =  e^{- \frac{1}{2}
  \left( \frac{\gamma}{t} \right)^2} e^{\pm \frac{i \gamma}{t} \hat{b}^+}
  e^{\pm \frac{i \gamma}{t} \hat{b}} \\
   = e^{- \frac{1}{2} \left( \frac{\gamma}{t} \right)^2} \left[ \exp
  \left( \mp \frac{i \gamma}{t} \sum_{j = 1}^{n_A} \sum_{N = 0}^{\infty}
  \sqrt{N + 1} \hat{c}_{i N }^+ \hat{c}_{i, N + 1} \right) \right]^+
   \exp \left( \pm \frac{i \gamma}{t} \sum_{j = 1}^{n_A} \sum_{N =
  0}^{\infty} \sqrt{N + 1} \hat{c}_{i N }^+ \hat{c}_{i, N + 1} \right)
  \nonumber\\
 =  e^{- \frac{1}{2} \left( \frac{\gamma}{t} \right)^2} \sum_{j =
  1}^{n_A} \sum_{k, M, N = 0}^{\infty} \frac{k!}{\sqrt{M!N!}}
  \left(\begin{array}{c}
    M\\
    k
  \end{array}\right) \left(\begin{array}{c}
    N\\
    k
  \end{array}\right) \left( \pm \frac{i \gamma}{t} \right)^{M + N - 2 k}
  \hat{c}_{j M}^+ \hat{c}_{j N} .
\end{multline}
Now we can rewrite the system Hamiltonian as
\begin{multline}
\label{eq-noninteracting-hamiltonian}
  \hat{H}  =  \hbar \omega \hat{b}^+ \hat{b} + \sum_{i = 1}^{n_A} U_i
  \hat{a}_{i }^+ \hat{a}_{i } + \sum_{\pm} \sum_{i = 1}^{n_A} t e^{\pm \frac{i
  \gamma}{t} (\hat{b} + \hat{b}^+)} \hat{a}_{i \pm 1 }^+ \hat{a}_{i }
  \\
 =  \sum_{i = 1}^{n_A} \sum_{N = 0}^{\infty} (U_i + N \hbar \omega)
  \hat{c}_{i N }^+ \hat{c}_{i N} +  \sum_{\pm} \sum_{i = 1}^{n_A} t e^{- \frac{1}{2} \left(
  \frac{\gamma}{t} \right)^2} \sum_{k, M, N = 0}^{\infty}
  \frac{k!}{\sqrt{M!N!}} \left(\begin{array}{c}
    M\\
    k
  \end{array}\right) \left(\begin{array}{c}
    N\\
    k
  \end{array}\right) \left( \pm \frac{i \gamma}{t} \right)^{M + N - 2 k}
  \hat{c}_{i \pm 1, M}^+ \hat{c}_{i N}, 
\end{multline}
and similarly for the lead terms. What is important is that the Hamiltonian
now has the same form as for noninteracting electrons,
\begin{equation}
  \hat{H} =  _{} \sum_{i j M N} h_{i M j N} \hat{c}_{i M}^+ \hat{c}_{j N} .
\end{equation}

We can also rewrite the charge density operator:
\begin{equation}
  \rho_i  = - e \hat{a}_{i }^+ \hat{a}_i = - e \sum_{N = 0}^{\infty}
  \hat{c}_{i N }^+ \hat{c}_{i N} 
\end{equation}
The current density operator can be deduced from the continuity equation.

Now it is obvious that the electric current can be calculated via the usual
Landauer-Buttiker formula applied to the 2d lattice described by
Hamiltonian (\ref{eq-noninteracting-hamiltonian}):
\begin{equation}
\label{eq-app-landauer}
  I  = \sum_{N_s} J_{N_d} \nonumber =  \sum_{N_s, N_d} \frac{e}{h} \int d E [n_s (E ; N_s) - n_d (E ; N_d)]
  T_{N_s N_d} (E), 
\end{equation}
where $E$ is the energy of the dressed fermion described by Hamiltonian (\ref{eq-noninteracting-hamiltonian}).

This derivation assumes there is only a single electron in the system, so that we can neglect e-e interactions. If we consider the multielectron case, but still neglect e-e interactions, the total current must be the sum of currents carried by each individual photon-dressed electron obeying Hamiltonian (\ref{eq-noninteracting-hamiltonian}). Expression (\ref{eq-app-landauer}) remains unchanged, only the occupation numbers now do not have
to sum up to unity.

Since electrons in the leads do not interact with photons, we can write the occupation numbers in the leads as products of electronic and photonic ones
\begin{equation}
  n_{s / d} (E ; N_{s / d})  =  n_{F \, s / d} (\varepsilon) \frac{e^{- \frac{N_{s / d}
  \hbar \omega}{k T}}}{1 + e^{- \frac{\hbar \omega}{k T}} + e^{- \frac{2 \hbar
  \omega}{k T}} + \cdots}    =  n_{F \, s / d} (E - N_{s / d} \hbar \omega) \frac{e^{- \frac{N_{s / d}
  \hbar \omega}{k T}}}{n_B (\hbar \omega) + 1} 
\end{equation}
and get

\begin{eqnarray}
  I & = & \frac{2 e}{h} \frac{1}{n_B (\hbar \omega) + 1} \sum_{N_s, N_d} \int
  d E \left[ n_{F\,s} (E - N_s \hbar \omega) e^{- \frac{N_s \hbar \omega}{k T}} -
  n_{F\, d} (E - N_d \hbar \omega) e^{- \frac{N_d \hbar \omega}{k T}} \right] T_{N_s
  N_d} (E) .
\end{eqnarray}

As the above derivation is meaningful only in the absence of the degeneracy, one may replace the Fermi distribution with its Boltzmann tail:
\begin{equation}
    n_{F\,s} (E - N_s \hbar \omega) e^{- \frac{N_s \hbar \omega}{k T}} \approx \exp\left\{ -\frac{E - N_s \hbar \omega - \mu_s}{kT} - \frac{N_s \hbar \omega}{k T} \right\} = \exp\left\{- \frac{E-\mu_s}{kT}\right\} \equiv f_s(E).
\end{equation}
As a result, the occupation numbers in the modified Landauer formula become the Boltzmann exponents of the total energy $E$, $f_{s/d}(E)$.

It is possible to show that the modified Landauer-type formula reduces to its original version (\ref{eq-Landauer-bare}) in the absence of electron-cavity coupling. In this limit, the photon number is unchanged during the electron passage, which formally reads as $\mathcal{T}_{NM} = \delta_{NM}\mathcal{T}_{NN}$. Further, the probability of tunneling depends only on the electron energy $\varepsilon = E - N\hbar\omega$, but not on the state of the cavity. The formal manifestation of that fact is represented by $\mathcal{T}_{NN}(E) = \mathcal{T}_{00}(E-N\hbar\omega)$. It is now convenient to shift the variable of energy integration in each $N$-th teem of the sum according to $\varepsilon = E-N\hbar\omega$. This shifts produces an extra multiplicative factor $f(E) = e^{-N\hbar\omega/kT}f(\varepsilon)$. We sum up these multiplicative factors $\sum_{N=0}^{\infty}e^{-N\hbar\omega/kT} = n_B(\omega)+1$. After this, we realize that all the information about the state of the cavity, contained in $n_B(\omega)$-functions, has dropped out, and we're left with an original Landauer formula (\ref{eq-Landauer-bare}).

\end{widetext}

\end{document}